# A General Approach to Modeling Covid-19

Raul Isea [1,*]

[1]Fundación IDEA, Hoyo de la Puerta, Baruta, Venezuela.

**Abstract**

The present work shows that it is possible to analytically solve a general model to explain the transmission dynamics of SARS-CoV-2. First, the within-host model is described, and later a between-host model, where the coupling between them is the viral load of SARS-CoV-2. The within-host model describes the equations involved in the life cycle of SARS-CoV-2, and also the immune response; while that the between-Host model analyzes the dynamics of virus spread from the original source of contagion associated with bats, subsequently transmitted to a host, and then reaching the reservoir (Huanan Seafood Wholesale Market in Wuhan ), until finally infecting the human population.

**Introduction**

The World Health Organization (WHO) reported 27 cases with a new severe respiratory syndrome of unknown etiology from Wuhan (Hubei province) in China [1]. Later identifies that it is a β-Coronavirus after sequencing the first days of January 2020 [2]. They originally called it 2019-nCOV, and after a month, it was renamed Severe Acute Respiratory Syndrome Coronavirus 2 (SARS-CoV-2), where the disease it produces is Covid-19. The WHO declared it an international public health emergency on January 30, 2020, and it was later declared a pandemic on March 11, 2020 [3].

So far the world has faced three coronavirus associated outbreaks. .The first was called SARS-CoV-1 [4], and it originated in the province of Guangdong Province in China: eight thousand cases were confirmed with a little less than eight hundred deaths in 2002 (affecting 29 countries until January 2004).The second incident was traced in Saudi Arabia, and for this reason, it was named MERS-CoV (Middle East Respiratory Syndrome Coronavirus), its first case was reported in June 2012 until November 2018, registered 2,494 cases after affecting 27 countries [5]. SARS-CoV-2 has infected more than 757 million people and 6 million deaths approximately until February 23, 2023, according to Johns Hopkins University (data available at coronavirus.jhu.edu).

It is currently a public health problem. Multiple vaccines have been developed to control the spread of the disease and to reduce the number of infections recorded daily in various parts of the world. However, it is necessary to do more studies, for example, on how to reduce the application of multiple doses regimens for a person, and so on. It can probably do identification of genomic patterns capable of generating an immune response in people derived from epitopes, as has been tested in other diseases [6,7,8], and later can be automated with the help of specialized workflows for automatic data handling [9].

Until this is achieved, it is necessary to develop mathematical models to design the public policies necessary to contain said infections based mainly on the implementation of social distancing measures, the use of face masks, quarantine programs and vaccination campaigns, etc.

In the scientific literature, there is a wide range of mathematical models that explain, for example, the contagion dynamics in various countries without a consensus on the prediction methodology (some examples in [10, 11]). Other works explain the transmission of viral particles in the environment [12], and also studies have also been carried out where they analyze the response of the immune system [13, 14]. However, it is necessary to integrate all these approaches on the same time scale. This process is known as immuno-epidemiological models of infectious disease systems [15] where they are described from a cellular level to the spread in the population. This type of model has been used in the dynamics of contagion in HIV-1 with a system of six differential equations [16, 17], as well as the paper of Murillo et al [18] by studying a multiscale model to explain the Influenza infection in 2013.

The present work presents a general model to explain the transmission dynamics of SARS-CoV-2, dividing the model in two sections. We first studied the within-host scale which we will divide into two different levels. The first consists of twelve compartments to describe the life cycle of SARS-CoV-2, and later the immune response. Finally, the between-host model consists of twelve compartments to describe the transmission of the virus from the source that gave rise to the virus (bats), the host (pangolins probably), the reservoir (Huanan Seafood Wholesale Market in Wuhan), until reaching the population human where we have added a compartment to explain the environmental viral load, a key element to integrate both scales.

**Mathematical model**

The mathematical model describes the contagion dynamics at two different sections scales, i.e., within-host model and between-host models. Each of them is described below.

*Within-host model*

The within-host model include the transmission dynamics from the life cycle of SARS-CoV-2, and subsequently the model that reproduces the immune response will be explained.

*SARS-CoV-2 life cycle*

The virus is a single-stranded RNA virus belonging to the Coronaviridae family, Coronavirinae subfamily [19]. The first reading frame, known as ORF1ab (the largest gene), occupies two-thirds of the virus sequence, approximately 28-32 kb. It is a 5'-capped and 3'-polyadenylated positive-sense single-strand RNA (+ssRNA), non-segmented and similar to the structure found in messenger RNA of eukaryotic cells. The replication of the virus begins when the S protein of SARSCoV-2 binds directly to the Angiotensin-Converting Enzyme 2 (ACE2) receptor. Once the virus has entered the host cell, it is released into the cytoplasm, starting the replication process that will give rise to non-structural proteins and also accessory proteins, with four structural proteins. The formation of RNA(-) and also replication and transcription of RNA subgenomics. These subgenomic RNAs(-) are transcribed into mRNAs(+) which encode the structural proteins S, M, E, N, and accessory proteins. During the replication process, the N protein of the virus binds to the genome, while the M protein associates with the membranes of the endoplasmic reticulum (ER). Finally, the virions are secreted by exocytosis [19].

Recently Isea and Mayo-García [20] have proposed the first analytical solution for the life cycle of SARS-CoV-2 according to the previous description. This model can be seen in figure 1, where it has been divided into four levels in different colors, where the first explains the cell entry, the second the genome transcription and replication, then the translation of structural and accessory proteins, and finally the assembly and eventual releases of new virions from the cell. The twelve variables that are going to describe the SARS-CoV-2 life cycle model are shown in Table 1, so the equations are simply (details in [20]):

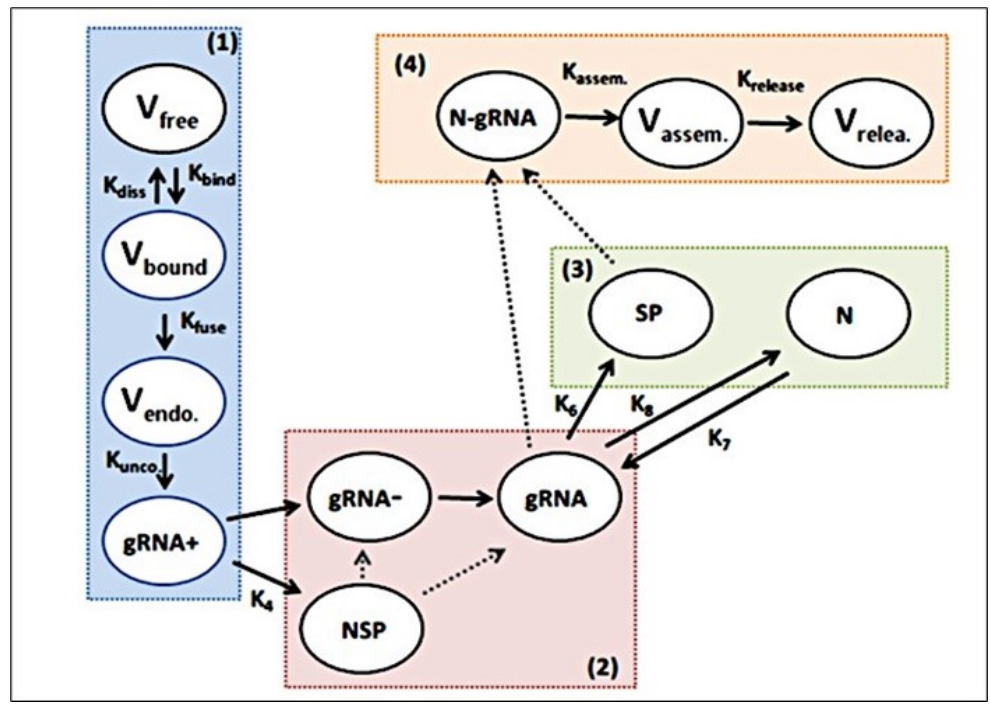

Figure 1. Diagram of the life cycle of SARS-CoV-2. The level that describes the cell entry is shown in blue color and indicated with number 1. The transcription and replication of the genome is in red color (number 2), while that the translation of structural and accessory proteins, and the assembly and release of virions are show in green and orange color, respectively. Only a few constants are shown in the life cycle dynamics and all the details are described in the text.

Here [$V_{release}$] is the concentration of viral particles that are released in the cell, and they are precisely the ones that will trigger the immune response in people, as will be explained in the next section.

$$\frac{d[V_{free}]}{dt} = K_{diss}[V_{bound}] - K_1[V_{free}]$$

$$\frac{d[V_{bound}]}{dt} = K_{bind}[V_{free}] - K_2[V_{bound}]$$

$$\frac{d[V_{endosome}]}{dt} = K_{fuse}[V_{bound}] - K_3[V_{endosome}]$$

$$\frac{d[gRNA_+]}{dt} = K_{uncoat}[V_{endosome}] - d_{gRNA}[gRNA_+]$$

$$\frac{d[NSP]}{dt} = K_4[gRNA_+] - d_{NSP}[NSP]$$

$$\frac{d[gRNA_-]}{dt} = K_{tr-}[NSP][gRNA_+] - d_{gRNA}[gRNA_-]$$

$$\frac{d[gRNA]}{dt} = K_{tr+}[NSP][gRNA_-] + K_8[N] - (K_{complex}[N] + d_{gRNA})[gRNA]$$

$$\frac{d[N]}{dt} = K_5[gRNA] - d_N[N]$$

$$\frac{d[SP]}{dt} = K_6[gRNA] - d_{SP}[SP]$$

$$\frac{d[N-gRNA]}{dt} = K_{complex}[N][gRNA] - (K_{assemble}[SP] + d_{N-gRNA})[N-gRNA]$$

$$\frac{d[V_{assemble}]}{dt} = K_{assemble}[SP][N-gRNA] - K_7[V_{assemble}]$$

$$\frac{d[V_{release}]}{dt} = K_{release}[V_{assemble}] - d[V_{release}]$$

Table 1: Description of the variables used to describe the life cycle of SARS-CoV-2 (details in [20]).

| Variable | Description | Variable | Description |
| --- | --- | --- | --- |
| $[V_{free}]$ | Free virions outside of cell | $[gRNA_-]$ | Negative sense genomic and subgenomic |
| $[V_{bound}]$ | Number of virions bound to ACE2 | $[gRNA]$ | Positive sense ge-nomic and subgenomics RNAs |
| $[V_{endosome}]$ | Number of virions in endosomes | $[N]$ | Concentration of $N$ proteins per virion |
| $[gRNA_+]$ | Number of ss-Positive sense genomic RNA | $[SP]$ | S+M+E per virion |
| $[NSP]$ | Abundance of nonstructural protein populations | $[N-gRNA]$ | Ribonucleocapsid molecules |
| $[V_{assemble}]$ | Assembled virions | $[V_{released}]$ | Released virions |

*Modelling the immune response*

The immune response that is triggered in people as a result of infection by SARSCoV-2 virions has been described in various models in the scientific literature [21, 22], but most posts solve it numerically. Furthermore, recent studies have associated a storm of cytokines generated by SARS-CoV-2 such that it leads lymphocytes to the lungs in search of infection, and thus maintains the replication and transmission of the virus [23].

Of all the models available in the scientific literature, we only consider the simplest of them, as shown in

$$\frac{dT}{dt} = \lambda_T - \beta VT - \mu T$$
$$\frac{dI}{dt} = \beta VT - \delta I$$
$$\frac{dV}{dt} = wI - cV - bAV$$
$$\frac{dA}{dt} = \alpha VA - \sigma A$$

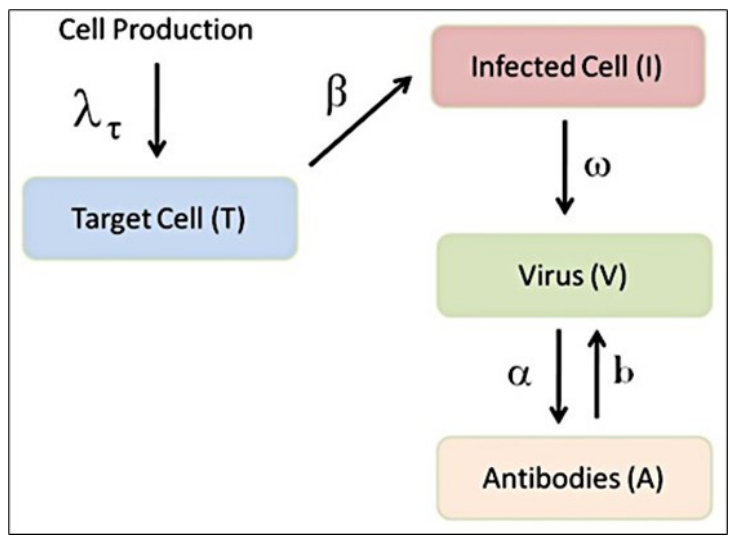

Figure 2: Diagram illustrating the immune response (see text for more details)

Figure 2, where the target cells (cells without contagion and represented by the letter T) are going to be infected (I) as a result of SARS-CoV-2 viral particles (V), so that an immune response product of antibodies is generated (A) (details in [22]). The equations that describe the above process are:

Where T, I, V, and A represent target cells, infected cells, viral particles, and antibodies, respectively. The definitions of parameters will be derived directly from the work of Danchin et al [24]. So both the life cycle of the virus and the dynamics of the immune response present the viral load, which is the link between them.

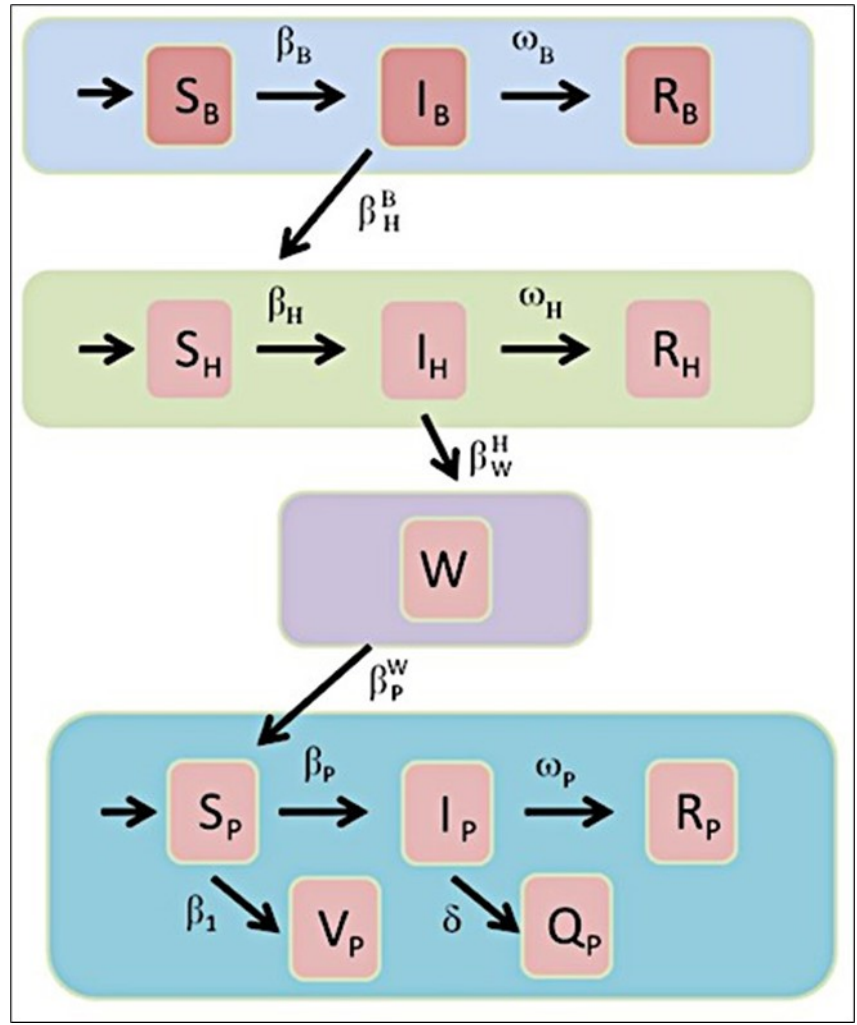

Figure 3: The Flowchart of the model proposed in this paper. The infection source is shown in blue color and it is a bats population, represented with B subscript. The host region is green and depicted with the H subscript (some bioinformatics studies indicated that could be pangolin population). The Reservoir (violet re- gion) was the Seafood Market in Wuhan. Finally, the human people are the last section and identified with the P subscript in cyan color.

$$\frac{dS_B}{dt} = \Lambda_B - \frac{\beta_B S_B I_B}{N_B} - m_b S_B$$

$$\frac{dI_B}{dt} = \frac{\beta_B S_B I_B}{N_B} - (\omega_B + m_B + \beta_H^B) I_B$$

$$\frac{dRB_B}{dt} = \omega_B I_B - m_B R_B$$

$$\frac{dS_H}{dt} = \Lambda_H + \beta_H^B I_B - \frac{\beta_H S_H I_H}{N_H} - m_H S_H$$

$$\frac{dI_H}{dt} = \frac{\beta_H S_H I_H}{N_H} - (\omega_H + m_H + \beta_\omega^H) I_H$$

$$\frac{dR_H}{dt} = \omega_H I_H - m_H R_H$$

$$\frac{dW}{dt} = \beta_W^H I_H - (\varepsilon + \beta_P^W) W$$

$$\frac{S_P}{dt} = \Lambda_P + \beta_P^W W - \frac{\beta_P S_P I_P}{N_P} - \beta_1 S_P V_P - m_P S_P$$

$$\frac{dI_P}{dt} = \frac{\beta_P S_P I_P}{N_P} + \beta_1 S_P V_P - (\omega_P + m_P + \delta) I_P$$

$$\frac{dQ_P}{dt} = \delta I_P - (m_P + \alpha_Q) Q_P$$

$$\frac{dR_P}{dt} = \omega_P I_P - m_P R_P + \alpha_Q Q_P$$

$$\frac{dV_P}{dt} = \beta_3 I_P V_P - \pi V_P$$

*Between-hosts model*

Figure 3 represents a model of the transmission dynamics from the possible origin of the virus associated with bats (denoted with the subscript B), then infecting an unknown host (usually associated with pangolins, abbreviated with the letter H). Later it reaches the reservoir (i.e., Wuhan market, W), until finally infecting the human population (P)

Therefore, this model that allow us to explain the transmission of the virus is given by the following differential equations:

We can observe this model is divided into twelve categories where the subscripts B, H, and P have been used to represent the bat, pangolin, and human, respectively. The Populations susceptible to contracting the virus are $S_B$, $S_H$, and $S_P$, and the population infected *are* $I_B$, $I_H$, and $I_P$. $R_B$, $R_H$, and $R_P$ represent the recovered bat, pangolin, and human populations, and $Q_P$ is the population that is in quarantine. The reservoir is denoted as W, and the last compartment is $V_P$, which represents the environmental viral load, respectively.

The constants $\Lambda_B$, $\Lambda_H$ and $\Lambda_P$ are the newborn bats, pangolins, and human, respectively. The contagion rate in bats, pangolins, and humans is given by the constants $\beta_B$, $\beta_H$, and $\beta_P$, respectively. Also, $m_B$, $m_H$, and $m_P$ represent the death rate, and $N_B$, $N_H$ and $N_P$ are the total populations of bats, pangolins, and human, respectively; $\omega_B$, $\omega_H$, and $\omega_P$ are the infectious period of bats, pangolins, and human, respectively. The infection rate between bats and pangolins is given by the variable $\beta^B_H$, where the superscript and subscript identify the beginning and end of the rate of infection, respectively; while the contagion rate from pangolins to the reservoir is given by $\beta^H_w$, and $\beta^W_P$ from reservoir to human. These values should be obtained by fitting the model with the collected data. Finally, $\varepsilon$ is the lifetime of the virus in the Reservoir. Given that each country maintains different policies to combat the virus, it is necessary to fit it according to infected data.

**Mathematical analysis and numerical calculations**

The resolution of the systems of equations have been validated in multiple scientific works [25, 26, 27, 28], so only the critical points and the eigenvalues of the system that allow us to elucidate the scenario to explain the dynamical of contagious of Covid-19.

*Analysis of the within-host model*

*SARS-CoV-2 life cycle*

The mathematical details have been recently published by Isea and Mayo-Garcia [20], where two critical points were derived as follows, noted with a letter each of the critical points obtained in each of the submodels descried above.

*The first critical point*

It is the trivial solution of the system and occurs when all variables are zero:

$$
\begin{aligned}
&[V_{free}]^* = 0, &&[V_{bound}]^* = 0, \\
&[V_{endosome}]^* = 0, &&[gRNA_+]^* = 0, \\
&[NSP]^* = 0, &&[gRNA_-]^* = 0, \\
&[gRNA]^* = 0, &&[N]^* = 0, \\
&[SP]^* = 0, &&[N-gRNA]^* = 0, \\
&[V_{assemble}]^* = 0, &&[V_{release}]^* = 0
\end{aligned}
$$

There is no contagion by the virus and therefore, there is no spread of the virus (i.e., disease-free equilibrium). The interesting of this critical point is that when calculating the Jacobian of the system and evaluating it at this critical point, the eigenvalues basically depends on the constants referring to the entry of the virus into the human body.

*Second critical point*

The second critical point corresponds to an endemic model where the replication and transcription process of the virus can be seen before being released:

$$
\begin{aligned}
&[V_{free}]^* = 0, &&[V_{bound}]^* = 0, \\
&[V_{endosome}]^* = 0, &&[gRNA_+]^* = 0, \\
&[NSP]^* = 0, &&[gRNA_-]^* = 0, \\
&[gRNA]^* = \frac{K_5 K_8 - d_N d_{gRNA}}{K_{complex} K_5}, &&[N]^* = \frac{K_5}{d_N}[gRNA]^*, \\
&[SP]^* = \frac{K_6[gRNA]^*}{d_{SP}}, \\
&[N-gRNA]^* = \left(\frac{K_{complex} d_{SP}[gRNA]^*}{d_{gRNA} d_{SP} + K_{assembled} K_6[gRNA]^*}\right)[N]^*, \\
&[V_{assemble}]^* = \left(\frac{K_{assembled} K_{complex} K_6[N]^*}{K_7(d_{SP} d_{gRNA} + K_{assembled} K_6[gRNA]^*)}\right)[gRNA]^{*2}, \\
&[V_{release}]^* = \left(\frac{K_{released}}{d_V}\right)[V_{assemble}]^*
\end{aligned}
$$

Table 2: Model parameters based on the Isea and Mayo-García's models [20].

| Parameter | Value | Parameter | Value |
|---|---|---|---|
| $K_{diss}$ | 0.61 | $d_{SP}$ | 0.04 |
| $K_{bind}$ | 12.00 | $d_{N-gRNA}$ | 0.12 |
| $K_{fuse}$ | 0.50 | $d_{endosome}$ | 0.06 |
| $K_{uncoat}$ | 0.50 | $K_1$ | 2.10 |
| $k_{complex}$ | 0.40 | $k_2$ | 2.99 |
| $K_{assemble}$ | 1.00 | $K_3$ | 0.20 |
| $K_{release}$ | 8.00 | $K_4$ | 0.19 |
| $d_{gRNA}$ | 0.20 | $K_5$ | 37.32 |
| $d_{NSP}$ | 0.07 | $K_6$ | 4.39 |
| $d_{gRNA}$ | 0.10 | $K_7$ | 8.01 |
| $d_N$ | 0.02 | $K_8$ | 0.02 |
| $K_{tr-}$ | 3.00 | $K_{tr+}$ | 1000.00 |

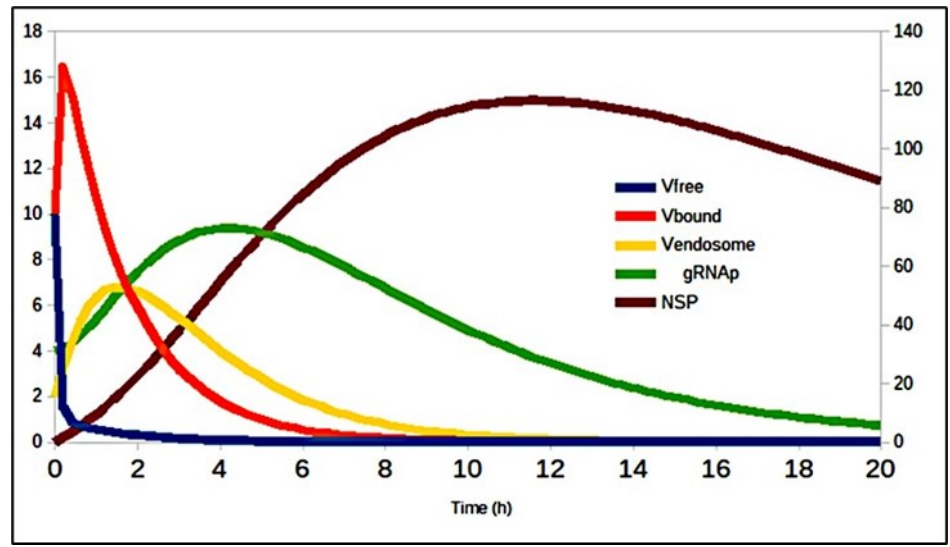

Figure 4: Evolution of virus transmission dynamics in the cell entry process (de- tails in [20]).

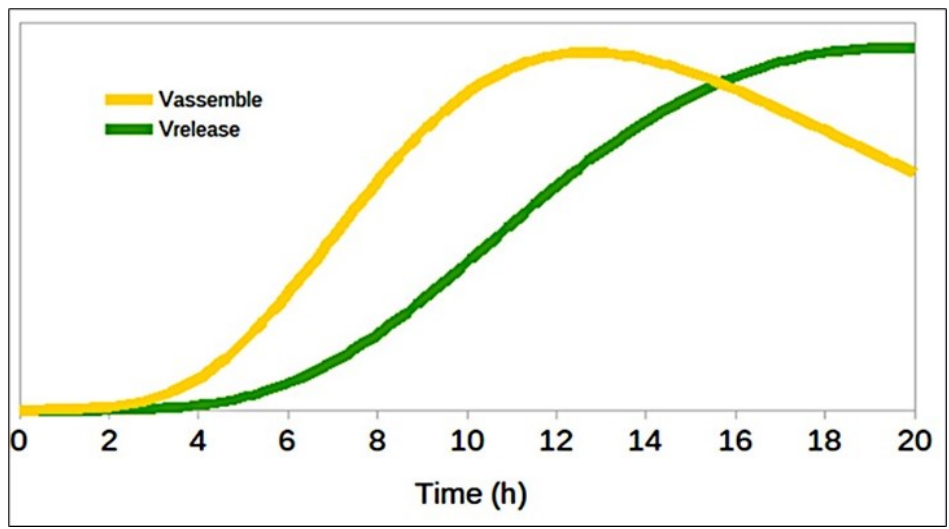

Figure 5: Number of assembled particles ([Vassembled]) and released ([Vreleased]).

Table 2 shows the values of the constants in this system of equations, and the graphic results are shown in Figures 4 and 5, where it can be seen both the contagion dynamics when it infects the cell (fig 4) as well as the number of particles that are assembled and released at the end of the SARS-CoV-2 life cycle (fig 5).

*Immune response*

The four differential equations that allow the generation of the immune response when infected by the virus are analyzed below. This model generates three critical points (marked with a different letter for easy identification) as explained below.

*The first critical point*

The first critical point corresponds when there is no presence of the virus in the body (i.e., disease-free equilibrium).

$$CP^1 : \left[ T^* = \frac{\lambda_T}{\mu}; \quad I^* = 0; \quad V^* = 0, \quad A^* = 0 \right]$$

Once the Jacobian evaluated at this critical point has been evaluated, two eigenvalues are obtained;

$$-\frac{\mu(\delta+c) + \sqrt{(c-\delta)^2\mu^2 + 4\beta\mu\lambda_T w}}{2\mu};$$

$$\frac{\sqrt{(c-\delta)^2\mu^2 + 4\beta\mu\lambda_T w} - \mu(\delta+c)}{2\mu}$$

When numerically evaluating these expressions according to the value of the different parameters (Table 3), we obtain -20.28 and +2.61, that is, the system is unstable (i.e., saddle point), so further studies are necessary to identify the stability conditions of the system as well as a stability analysis in said system.

*Second critical point*

The second value corresponds to the case when the infection process is being generated, without generating an antibody response, described as follows:

$$CP^2 : \left[ T^* = \frac{c\delta}{\beta w}; \quad I^* = \frac{\lambda_T - \mu T^*}{\delta}; \quad V^* = \frac{w}{c} I^*; \quad A^* = 0 \right]$$

The eigenvalues in the system are:

$$-\frac{c\delta(c+\delta) + \beta w \lambda_T}{3c\delta}; \quad -\frac{c\delta(\beta\delta + \alpha\mu) - \alpha\beta w \lambda_T}{c\beta\delta}$$

When evaluating these eigenvalues, the following values are obtained: -1.86E+7, and 15.53, so it is again unstable at this critical point (saddle point).

*Third critical point*

The last critical point is the one where the immune response is activated by the presence of antibodies, obtaining:

$$CP^3 : \left[ T^* = \frac{\alpha \lambda_T}{\alpha \mu + \beta \sigma}; I^* = \frac{\beta \sigma \lambda_T}{\delta(\alpha \mu + \beta \sigma)}; V^* = \frac{\sigma}{\alpha}; A^* = \frac{\beta w T^* - c\delta}{b\delta} \right]$$

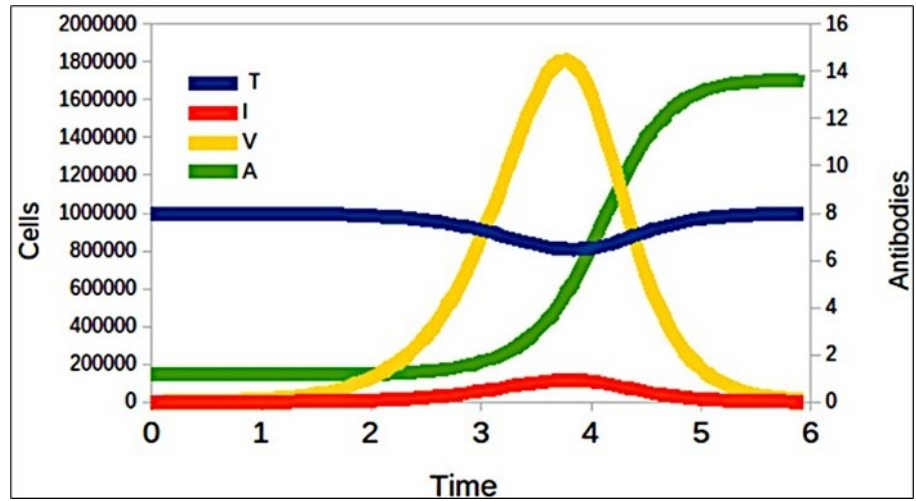

Figure 6: Immune response generated by target cells (T), cells infected (I) by SARS-CoV-2 virions (V), and antibodies generated and denoted as A.

Table 3: Parameters defined in the dynamics of viral infection according to the work of Danchin et al [24].

| Parameter | value |
|---|---|
| $\mu$ | 9.66 |
| $\lambda_T$ | $9.66 \times 10^6$ |
| $\beta$ | $1.28 \times 10^{-6}$ |
| $\delta$ | 16.22 |
| c | 1.45 |
| b | 0.52 |
| $\sigma$ | 0.02 |
| w | 59.74 |
| $\alpha$ | $9.15 \times 10^{-7}$ |

Unfortunately, it has not been an easy task to derive the expression of the eigenvalues for this case.

It only remains to visualize how the immune response varies against the pres- ence of the virus as identified in Figure 6, according to the values of the parameters indicated in Table 3 (these values were obtained from the work of Danchin et al [24]).

*Between-hosts model*

The analytical resolution of the system of equations generates twelve different scenarios represented by the twelve critical points, where the following critical point notation (where *i* runs from 1 to 12) has been used to identify each of the scenarios derived in the work.

*First critical point*

The first critical point (CP$^1$) is the trivial solution of the system, that is, when there are no infections in any of the populations:

$$CP^1 : \begin{bmatrix} S_B^* = \frac{\Lambda_B}{m_B}, & I_B^* = 0, \\ R_B^* = 0, & S_H^* = \frac{\Lambda_H}{m_H}, \\ I_H^* = 0, & R_H^* = 0, \\ W^* = 0, & S_P^* = \frac{\Lambda_P}{m_P}, \\ I_P^* = 0, & Q_P^* = 0 \\ R_P^* = 0, & V_P^* = 0 \end{bmatrix}$$

This result indicates the population that may be susceptible to contracting the virus are $\Lambda_B/m_B$, $\Lambda_H/m_H$ and $\Lambda_P/m_P$ corresponding to the populations of bats, pangolins and humans, respectively.

The next step is to determine the equilibrium conditions. The following notation has been used to describe the eigenvalues: the superscript indicates the critical point, while the subscript is the number of different eigenvalues obtained. For example, $\lambda_3^1$ corresponds to the third eigenvalue of the first critical point. By doing this, three different eigenvalues are obtained:

$$\lambda_1^1 = -\frac{\Delta_H}{N_H m_H}; \quad \lambda_2^1 = -\frac{\Delta_B}{N_B m_B}; \quad \lambda_3^1 = -\frac{\Delta_P}{N_P m_P}$$

where three new variables ($\Delta_H$, $\Delta_B$ and $\Delta_P$) have been introduced indicated in Table 4 to simplify the equation in this paper. These solutions reveal that the different populations depend exclusively on each population and are not conditioned on each other.

*Second critical point*

The second critical point (CP$^2$) is when only a fraction of susceptible human population without presence of the virus in the environment (W$^*$ = 0, V$_P^*$ = 0) is considered:

$$CP^2 : \begin{bmatrix} S_B^* = \frac{\Lambda_B}{m_B}, & I_B^* = 0, \\ R_B^* = 0, & S_H^* = \frac{\Lambda_H}{m_H}, \\ I_H^* = 0, & R_H^* = 0, \\ W^* = 0, & S_P^* = \frac{N_P P_0}{\beta_P}, \\ I_P^* = \frac{\Lambda_P - m_P S_P^*}{P_0}, & Q_P^* = \frac{\delta I_P^*}{Q_0} \\ R_P^* = \left(\frac{w_P Q_0 + \delta \alpha_Q}{m_P Q_0}\right) I_P^*, & V_P^* = 0 \end{bmatrix}$$

This result reveals that it is possible to circulate the virus in the population of people who may be infected without the virus being present in the environment, infecting other people. By repeating the procedure described above, five eigenvalues are obtained:

$$\lambda_1^2 = -\frac{\Delta_B}{N_B m_B}; \quad \lambda_2^2 = -\frac{\Delta_H}{N_H m_H}$$

$$\lambda_3^2 = -\frac{\beta_P \Lambda_P + \sqrt{\beta_P^2 \Lambda_P^2 + 4 N_P P_0 \Delta_P}}{2 N_P P_0}; \quad \lambda_4^2 = \frac{\sqrt{\beta_P^2 \Lambda_P^2 + 4 N_P P_0^2 \Delta_P} - \beta_P \Lambda_P}{2 N_P P_0}$$

$$\lambda_5^2 = -\frac{\beta_3 P_0 \pi + \beta_B \Delta_P}{\beta_P P_0}$$

The first two eigenvalues correspond to the equilibrium condition when the virus is not present in bats or in the host, while the last three correspond to con- ditions in the human population. These values must be adjusted with the records of contagions made in the world, as will be explained in future works.

Table 4: Definitions made to simplify mathematical expressions obtained at work.

| |
|---|
| $B_0 = m_B + w_B + \beta^B{}_H$ |
| $H_0 = m_H + w_H + \beta^H{}_W$ |
| $P_0 = m_P + w_P$ |
| $\Delta_B = N_B m_B B_1 - \beta_B \Lambda_B$ |
| $\Delta_H = N_H m_h H_1 - \beta_H \Lambda_H$ |
| $\Delta_P = N_P m_P P_1 - \beta_P \Lambda_P$ |
| $E_0 = \varepsilon + \beta^W{}_P$ |

*Third critical point*

The third critical point ($CP^3$) relates to the scenario when the bat population is infected, and represents the origin of the infection in the human population, so that the host can simply indicate the fraction of people susceptible to such disease, and described by the following equations:

$$CP^3 : \begin{bmatrix} S_B^* = \frac{N_B B_0}{\beta_B}, & I_B^* = \frac{\Lambda_B - m_B S_B^*}{B_0}, \\ R_B^* = \left(\frac{w_B}{m_B}\right) I_B^*, & S_H^* = \frac{\Lambda_H + \beta_H^B I_B^*}{m_H}, \\ I_H^* = 0, & R_H^* = 0, \\ W^* = 0, & S_P^* = \frac{\Lambda_P}{m_P}, \\ I_P^* = 0, & Q_P^* = 0 \\ R_P^* = 0, & V_P^* = 0 \end{bmatrix}$$

It should be noted that the host population that can be infected by the virus depends directly on the rate of infection of the bat. The eigenvalues obtained are:

$$\lambda_1^3 = -\frac{\beta_B \Lambda_B + \sqrt{\beta_B^2 \Lambda_B^2 + 4 N_B B_0^2 \Delta_B}}{2 N_B B_0}; \quad \lambda_2^3 = \frac{\sqrt{\beta_B^2 \Lambda_B^2 + 4 N_B B_0^2 \Delta_B} - \beta_B \Lambda_B}{2 N_B B_0}$$

$$\lambda_3^3 = -\frac{\Delta_P}{N_P m_P}; \quad \lambda_4^3 = -\frac{\beta_H \Delta_B \beta_H^B + \beta_B B_0 \Delta_H}{N_H m_H \beta_B B_0}$$

The last eigenvalues are a combination of conditions between the bat and host populations, and the equilibrium conditions depend on the values $\Delta_B$ and $\Delta_H$. Actually, these values should not be easy to obtain because there are still many questions about how the virus affects the human population.

*Fourth critical point*

The fourth critical point ($CP^4$) represents the scenario when the bat population and the human population are infected, without the host or the environment presenting the virus.

$$CP^4 : \begin{bmatrix} S_B^* = \frac{N_B B_0}{\beta_B}, & I_B^* = \frac{\Lambda_B - m_B S_B^*}{B_0}, \\ R_B^* = \left(\frac{w_B}{m_B}\right) I_B^*, & S_H^* = \frac{\Lambda_H - \beta_H^B I_B^*}{m_H}, \\ I_H^* = 0, & R_H^* = 0, \\ W^* = 0, & S_P^* = \frac{N_P P_0}{\beta_P}, \\ I_P^* = \frac{\Lambda_P - m_P S_P^*}{P_0}, & Q_P^* = \frac{\delta I_P^*}{Q_0}, \\ R_P^* = \left(\frac{w_P Q_0 + \delta \alpha_Q}{m_P Q_0}\right) I_P^*, & V_P^* = 0 \end{bmatrix}$$

Where the new variables indicated in this critical point are listed in Table 4. The six eigenvalues in this scenario are:

$$\lambda_1^4 = -\frac{\beta_B \Lambda_B + \sqrt{\beta_B^2 \Lambda_B^2 + 4 N_B B_0^2 \Lambda_B}}{2 N_B B_0}; \quad \lambda_2^4 = \frac{\sqrt{\beta_B^2 \Lambda_B^2 + 4 N_B B_0^2 \Lambda_B} - \beta_B \Lambda_B}{2 N_B B_0}$$

$$\lambda_3^4 = -\frac{\beta_H \Delta_B \beta_H^B + \beta_B B_0 \Delta_H}{N_H m_H \beta_B B_0}; \quad \lambda_4^4 = -\frac{\beta_B \Delta_P + \beta_P P_0 \pi}{\beta_P P_0}$$

$$\lambda_5^4 = -\frac{\beta_P \Lambda_P + \sqrt{\beta_P^2 \Lambda_P^2 + 4 N_P P_0^2 \Delta_P}}{2 N_P P_0}; \quad \lambda_6^4 = \frac{\sqrt{\beta_P^2 \Lambda_P^2 + 4 N_P P_0^2 \Delta_P} - \beta_P \Lambda_P}{2 N_P P_0}$$

Expressions of the combinations between species are already beginning to be obtained, which complicates the obtaining of said equations.

*Fifth critical point*

The fifth critical point ($CP^5$) corresponds to the case when the population of pangolins is infected and the virus is also in the Reservoir, so a part of the human population is susceptible to contracting this disease, that is:

$$CP^5 : \begin{bmatrix} S_B^* = \frac{\Lambda_B}{m_B}, & I_B^* = 0, \\ R_B^* = 0, & S_H^* = \frac{N_H H_0}{\beta_H}, \\ I_H^* = \frac{\Lambda_H - m_H S_H^*}{H_0}, & R_H^* = \left(\frac{w_H}{m_H}\right) I_H^*, \\ W^* = \frac{(\beta_H \Lambda_H - H_0 N_H m_H)\beta_W^H}{\beta_H H_0 E_0}, & S_P^* = \frac{\Lambda_P - \beta_P^W W^*}{\beta_P}, \\ I_P^* = 0, & Q_P^* = 0 \\ R_P^* = 0, & V_P^* = 0 \end{bmatrix}$$

This solution reveals how the environment ($W^*$) plays an important role in the human population that can contract this disease. The four eigenvalues are:

$$\lambda_1^5 = -\frac{\beta_H \Lambda_H + \sqrt{\beta_H^2 \Lambda_H^2 + 4N_H H_0^2 \Delta_H}}{2N_H H_0}; \quad \lambda_2^5 = \frac{\sqrt{\beta_H^2 \Lambda_H^2 + 4N_H H_0^2 \Delta_H} - \beta_H \Lambda_H}{2N_H H_0}$$

$$\lambda_3^5 = -\frac{\Delta_B}{N_B m_B}; \quad \lambda_4^5 = -\frac{(\beta_H \Lambda_H - H_0 N_H m_H) H_0 E_0 \beta_H + \beta_P \beta_W^H \beta_P^W \Delta_H}{N_P m_P \beta_H H_0}$$

As can be seen above, the first two eigenvalues depend exclusively on the parameters of the pangolins, then on the bats, and finally the last is a combination of the parameters between the host and human people.

*Sixth critical point*

The sixth critical point ($CP^6$) corresponds to the case when bat populations and humans are infected, that is:

$$CP^6 : \begin{bmatrix} S_B^* = \frac{\Lambda_B}{m_B}, & I_B^* = 0, \\ R_B^* = 0, & S_H^* = \frac{N_H H_0}{\beta_H}, \\ I_H^* = \frac{\Lambda_H - m_H S_H^*}{H_0}, & R_H^* = \left(\frac{w_H}{m_H}\right) I_H^*, \\ W^* = \frac{\beta_W^H I_H^*}{E_0}, & S_P^* = \frac{N_P P_0}{\beta_P}, \\ I_P^* = \frac{\beta_W^H \beta_P \Delta_H + H_0 \beta_H E_0 \overline{\Delta_P}}{E_0 P_0 H_0 \beta_P}, & Q_P^* = \frac{\delta I_P^*}{Q_0} \\ R_P^* = \left(\frac{w_P Q_0 + \delta \alpha_Q}{m_P Q_0}\right) I_P^*, & V_P^* = 0 \end{bmatrix}$$

The eigenvalues from this critical point have not been easy to obtain.

*Seventh critical point*

The seventh critical point ($CP^7$) corresponds to the case when the virus is present in bats, pangolins, and in the reservoir.

$$CP^7 : \begin{bmatrix} S_B^* = \frac{N_B B_0}{\beta_B}, & I_B^* = \frac{\Lambda_B - m_B S_B^*}{B_0}, \\ R_B^* = \left(\frac{w_B}{m_B}\right) I_B^*, & S_H^* = \frac{N_H H_0}{\beta_H}, \\ I_H^* = \frac{\Lambda_H - m_H S_H^* - \beta_H^B I_B^*}{H_0}, & R_H^* = \left(\frac{w_H}{m_H}\right) I_H^*, \\ W^* = \frac{\beta_W^H I_H^*}{E_0}, & S_P^* = \frac{\Lambda_P + \beta_P^W W^*}{m_P}, \\ I_P^* = 0, & Q_P^* = 0 \\ R_P^* = 0, & V_P^* = 0 \end{bmatrix}$$

It should be noted that the human population that is susceptible to contracting the virus depends significantly on the reservoir. Only we can calculate, the eigenvalues corresponding to the bat population.

$$\lambda_1^7 = -\frac{\beta_B \Lambda_B + \sqrt{\beta_B^2 \Lambda_B^2 + 4N_B B_0^2 \Delta_B}}{2N_B B_0};$$

$$\lambda_2^7 = \frac{\sqrt{\beta_B^2 \Lambda_B^2 + 4N_B B_0^2 \Delta_B} - \beta_B \Lambda_B}{2N_B B_0}$$

*Eighth critical point*

The last critical point ($CP^8$) corresponds to the case when the three populations are infected by Covid-19, without being present at $V^*_P = 0$, and this result should be analyzed in more detail in the future .

$$CP^8 : \begin{bmatrix} S_B^* = \frac{N_B B_0}{\beta_B}, & I_B^* = \frac{\Lambda_B - m_B S_B^*}{B_0}, \\ R_B^* = \left(\frac{w_B}{m_B}\right) I_B^*, & S_H^* = \frac{N_H H_0}{\beta_H}, \\ I_H^* = \frac{\Lambda_H - m_H S_H^* - \beta_H^B I_B^*}{H_0}, & R_H^* = \left(\frac{w_H}{m_H}\right) I_H^*, \\ W^* = \frac{\beta_W^H I_H^*}{E_0}, & S_P^* = \frac{N_P P_0}{\beta_P}, \\ I_P^* = \frac{\Lambda_P - m_P S_P^* + \beta_P^W W^*}{P_0}, & Q_P^* = \frac{\delta I_P^*}{Q_0} \\ R_P^* = \left(\frac{w_P Q_0 + \delta \alpha_Q}{m_P Q_0}\right) I_P^*, & V_P^* = 0 \end{bmatrix}$$

This critical point is when everyone is infected, and unfortunately it has not been possible to determine the eigenvalues for this system.

*Ninth critical point*

This critical point ($CP^9$) corresponds to the case when the human population is present the virus, where $V_P^* \neq 0$, and should be the scenario capable of describing the current transmission of the virus in different parts of the world.

$$CP^9 : \begin{bmatrix} S_B^* = \frac{\Lambda_B}{m_B}, & I_B^* = 0, \\ R_B^* = 0, & S_H^* = \frac{\Lambda_H}{m_H}, \\ I_H^* = 0, & R_H^* = 0, \\ W^* = 0, & S_P^* = \frac{\beta_3 \Lambda_P - \pi P_0}{\beta_3 m_P}, \\ I_P^* = \frac{\pi}{\beta_3}, & Q_P^* = \frac{\delta I_P^*}{Q_0} \\ R_P^* = \left(\frac{w_P Q_0 + \delta \alpha_Q}{m_P Q_0}\right) I_P^*, & V_P^* = \left(\frac{\beta_3 \Delta_P + \beta_P P_0 \pi}{\beta_1 N_P V_0}\right) I_P^* \end{bmatrix}$$

This scenario tells us how the virus is transmitted in the human population once it has been infected by the environment.

*Tenth critical point*

The tenth one corresponds to the scenario where the bats, the reservoir, and the human population are infected such that $V_P^* \neq 0$

$$CP^{10} : \begin{bmatrix} S_B^* = \frac{N_B B_0}{\beta_B}, & I_B^* = \frac{\Lambda_B - m_B S_B^*}{B_0}, \\ R_B^* = \left(\frac{w_B}{m_B}\right) I_B^*, & S_H^* = \frac{\Lambda_H - \beta_H^B I_B^*}{m_H}, \\ I_H^* = 0, & R_H^* = 0, \\ W^* = 0, & S_P^* = \frac{\beta_3 \Lambda_P - \pi P_0}{\beta_3 m_P}, \\ I_P^* = \frac{\pi}{\beta_3}, & Q_P^* = \frac{\delta I_P^*}{Q_0} \\ R_P^* = \left(\frac{w_P Q_0 + \delta \alpha_Q}{m_P Q_0}\right) I_P^*, & V_P^* = \left(\frac{\beta_3 \Delta_P + \beta_P P_0 \pi}{\beta_1 N_P V_0}\right) I_P^* \end{bmatrix}$$

Unfortunately it has not been easy to drive the eigenvalues.

*Eleventh critical point*

The penultimate critical point corresponds to the case when the virus is present in the host, the environment, and the human population.

$$CP^{11} : \begin{bmatrix} S_B^* = \frac{\Lambda_B}{m_B}, & I_B^* = 0, \\ R_B^* = 0, & S_H^* = \frac{N_H H_0}{\beta_H}, \\ I_H^* = \frac{\Lambda_H - m_H S_H^*}{H_0}, & R_H^* = \left(\frac{w_H}{m_H}\right) I_H^*, \\ W^* = \frac{\beta_W^H I_H^*}{E_0}, & S_P^* = \frac{\beta_3 \Lambda_P - \beta_3 \beta_P^W W^* - \pi P_0}{\beta_3 m_P}, \\ I_P^* = \frac{\pi}{\beta_3}, & Q_P^* = \frac{\delta I_P^*}{Q_0}, \\ R_P^* = \left(\frac{w_P Q_0 + \delta \alpha_Q}{m_P Q_0}\right) I_P^*, & V_P^* = \frac{\pi (P_0 N_P + \beta_P S_P^*)}{\beta_1 \beta_3 N_P S_P^*} \end{bmatrix}$$

Unfortunately it has not been possible to drive the eigenvalues.

*Last critical point*

The last critical point is when the virus circulates and is infecting all species. It is the most general possible model that can explain the spread of the virus from its origin.

$$CP^{12} : \begin{bmatrix} S_B^* = \frac{N_B B_0}{\beta_B}, & I_B^* = \frac{\Lambda_B - m_B S_B^*}{B_0}, \\ R_B^* = \left(\frac{w_B}{m_B}\right) I_B^*, & S_H^* = \frac{N_H H_0}{\beta_H}, \\ I_H^* = \frac{\Lambda_H - m_H S_H^* - \beta_H^B I_B^*}{H_0}, & R_H^* = \left(\frac{w_H}{m_H}\right) I_H^*, \\ W^* = \frac{\beta_W^H I_H^*}{E_0}, & S_P^* = \frac{\beta_3 \Lambda_P - \beta_3 \beta_P^W W^* - \pi P_0}{\beta_3 m_P}, \\ I_P^* = \frac{\pi}{\beta_3}, & Q_P^* = \frac{\delta I_P^*}{Q_0}, \\ R_P^* = \left(\frac{w_P Q_0 + \delta \alpha_Q}{m_P Q_0}\right) I_P^*, & V_P^* = \frac{\pi (P_0 N_P + \beta_P S_P^*)}{\beta_1 \beta_3 N_P S_P^*} \end{bmatrix}$$

This scenario is one where the virus is transmitted between all species. Unfortunately it has not been easy to derive the eigenvalues.

It only remains to indicate, as an example, some of the values of the constants used in the last equations. These values can be seen in a study carried out by Isea [28] after adjusting the data according to the contagion data present in Australia, using a least squares adjustment, where there is infection in the population of neither bats nor pangolins. The values obtained were: $\beta_P^W = 1.708$, $\beta_P = 1.215$, $\omega_P = 0.001$, $\Lambda_P = 81.212$, $\mu = 1.100$, $\varepsilon = 0.001$ (setting the $m_P$ value to 0.0000456). The value of $V_P$ must be adjusted with experimental data under a range of hypotheses that will be presented in a future work.

**Conclusion**

In the present work, a mathematical model has been proposed to study the transmission dynamics of Covid-19 from two different scales where the viral load of the virus is the key to unifying both these time scales. The advantage of this model is that it is possible to focus on a particular study scheme derived directly from the equilibrium conditions of the system, that is, there is a mathematical basis that justifies the existence of said scenario based on the critical points derived in the work. It should be pinpointed that all the models have been described with a deterministic model and the conditions to reach equilibrium have been obtained in most of them. In the next publication, the between-host model will be analyzed considering the sensitivity analysis and also the measures of social distancing and how quarantine affects the dynamics of contagion will be described.

**Acknowledgments**

**Conflict of interest**

The author declares that there is no conflict of interest.

**Data Availability**

Data available within the article.